# X-RAY BINARIES AND THEIR DESCENDANTS: BINARY RADIO PULSARS; EVIDENCE FOR THREE CLASSES OF NEUTRON STARS?

E.P.J. van den Heuvel

*Astronomical Institute "Anton Pannekoek" and Center for High Energy Astrophysics,
University of Amsterdam, The Netherlands, edvdh@science.uva.nl*

## ABSTRACT

An important recent discovery by Pfahl et al.(2002) is that there are two classes of Be X-ray binaries: one with orbits of small eccentricity (< 0.25), in which the neutron stars received hardly any kick velocity at birth and a class with substantial orbital eccentricities, in which the neutron stars received a kick velocity of order 100 km/s at birth. Also many of the double neutron stars (5 out of 7) have a low orbital eccentricity (0.09 to 0.27), which indicates that their second-born neutron stars received hardly any velocity kick at birth. These second-born neutron stars tend to have low masses (1.25 to 1.30 $M_\odot$). It is proposed that the low-mass, low-kick neutron stars formed by the electron-capture collapse of degenerate O-Ne-Mg cores of stars with initial masses below about 12-14 $M_\odot$, while the high-kick neutron stars originated from the photo-disintegration collapse of the iron cores of stars which started out with masses larger than this limit. The latter group may be further subdivided into two classes of different mass.

## 1. INTRODUCTION

There are many types of binaries in which one or both components are compact objects. In this presentation I restrict myself to recent results regarding to three topics:
(i) The formation and evolution of B-emission X-ray binaries and their descendants, the close double neutron stars. I particularly focus on the role of kicks imparted to neutron stars at their birth and on what can be deduced from the B-emission binaries and double neutron stars regarding the nature of these kicks and the formation processes of neutron stars.
(ii) Black hole X-ray binaries and the evidence for velocity kicks imparted to stellar black holes at their birth.
(iii) The accreting millisecond X-ray pulsars and the relation between the spin of the neutron stars and the kilohertz Quasi-Periodic Oscillations (QPO) in these systems. In view of space I will in these proceedings only focus on the first topic. For the second topic I refer to the recent paper by Jonker and Nelemans (2004) and references therein, which provide evidence indicating that black holes do receive a velocity kick at birth.
For the third topic I refer to the recent papers by Wijnands et al. (2003), and by Chakrabarty et al. (2003). There are now five millisecond X-ray pulsars, all have companions of very small mass and in these systems the difference between the kilohertz QPO frequencies is either the spin frequency of the neutron star or half the spin frequency of the neutron star, indicating that the kilohertz QPO phenomenon is related to the neutron star spin, as had long been suspected but so far could not be proven.

## 2. B-EMISSION X-RAY BINARIES, DOUBLE NEUTRON STARS, NEUTRON STAR KICKS AND DIFFERENT FORMATION MECHANISMS OF NEUTRON STARS

### 2.1. The B-emission X-ray binaries and Be stars with radio pulsar companions.

The B-emission X-ray binaries form the by far largest class of High Mass X-ray Binaries (HMXB) with at present over 80 identified members in our Galaxy and the Magelllanic Clouds. In contrast, the "other" class, of the so-called "standard" HMXBs like Cen X-3, Vela X-1 and Cyg X-1, has only about a dozen known members. In the latter class the orbital periods are typically short (less than about 10 days) and the companions are very massive (18 to over 40 $M_\odot$). On the other hand, in the B-emission X-ray binaries the orbital periods range from about 15 days to several years and the companions are of lower mass, typically between 8 and about 20 solar masses. While the "standard" systems are persistent X-ray sources, most of the B-emission X-ray binaries are transients which appear as X-ray sources only when their B-emission companion goes through a mass ejection phase. B-emission stars are very rapid rotators and their outburst phases occur erratically, and are often separated by many years of quiescence (see the reviews by Rappaport and van den Heuvel, 1982 and by van den Heuvel and Rappaport 1987).
The mass ejection from the Be star is somehow connected with its rapid rotation and occurs in its equatorial regions, leading to the formation of an equatorial gas disk around the star. The Be/X-ray Binaries often have quite eccentric orbits and during an outburst phase of the Be star the neutron star accretes matter every time it passes the periastron of its orbit and crosses the equatorial gas disk of the Be star. This leads then to periodic transient outbursts which repeat with the orbital period of the neutron star (see figure 1).

Recently it was pointed out by Pfahl et al. (2002) that apart from the Be/X-ray binaries with substantial orbital eccentricities there appears to be a subclass of the Be/X-ray binaries in which the systems have low orbital eccentricities (< 0.25). These authors found that 6 out of the 13 systems with known eccentricities have such low eccentricities. On the other hand, the other Be/X-ray binaries have relatively high orbital eccentricities, between 0.3 and 0.6. In addition, four binary radio pulsars have been discovered in which the companion of the radio pulsar is an early B-star of type similar to the Be companions in the Be/X-ray binaries. In 3 of these 4 cases the B-type companion is indeed a B-emission star. Table 1 lists the characteristic parameters of these four systems. The table shows that these radio pulsars tend to be very young (ages ranging from $10^5$ to $10^6$ years) and that the orbital eccentricities in all these four systems are very large, ranging from 0.58 to 0.87 (see Stairs 2004). These systems are expected to later on, when the rotation of their pulsars has sufficiently slowed down, evolve into Be/X-ray binaries. The fact that all these systems have very large orbital eccentricities may well be a selection effect, as the radio pulsars are observable only when they are far away from the circumstellar matter around the B-star, since otherwise the radio pulsar signals will be completely washed out by dispersion. So, only if the neutron star is able to reach a large distance from its companion may it be observable as a radio pulsar. This will require a wide orbit and a large orbital eccentricity. The only exception is the system PSR J0045-7319 in the Small Magellanic Cloud (Kaspi 1996), in which the B-star is not an emission line star (so: there is no gas around the star). This star is rotating in a direction opposite to the orbital motion of the neutron star (Lai 1996). The latter is a clear sign that the neutron star received such a large kick at its birth that the direction of its orbital motion was reversed. (see further below).

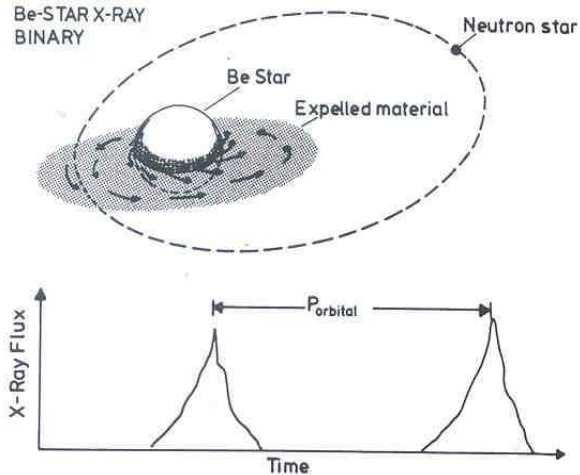

*Fig.1 Schematic model of a B-emission X-ray binary. When the rapidly rotating B star ejects an equatorial disk of gas it goes through an emission-line phase and the neutron star goes through an X-ray phase. During this phase it turns on periodically, each time when the neutron star in its eccentric orbit passes through the disk. Mass-ejection phases of the Be star occur at irregular times which may be separated by many years of quiescence. Recently it was found that a substantial fraction of the B-emission X-ray binaries has rather small orbital eccentricities (< 0.25; Pfahl et al.2002).*

*Table 1 Young radio pulsars with B-star companions*

| Name | $P_{orb}$ (d) | e | comp type | $P_{pulse}$ (ms) | Age $P/2\dot{P}$ (yr) | $M_B$ ($M_\odot$) | Ref. |
|---|---|---|---|---|---|---|---|
| J0045-7319 | 51.17 | 0.81 | B1V | 926.3 | ~ $10^6$ | ~10 | (1) |
| B1259-63 | 1236.7 | 0.87 | B0e | 47.8 | ~ $10^5$ | ~12 | (1) |
| J1638-4715 | ~ 1800 | 0.808 | Be ? | 764 | ? | ? | (2) |
| J1740-3052 | 231 | 0.579 | Be | 570.03 | $3.6 \times 10^5$ | ≥11 | (1) |

References: (1) Stairs (2004) ; (2) Lyne (2004)

**2.2. Evolutionary history of the B-emission X-ray binaries: need for kicks to obtain the eccentricities of the high-eccentricity systems.**

Figure 2 depicts the standard evolutionary model for the formation of the B-emission X-ray binaries developed by van den Heuvel (1983) and Habets (1986a,b). The systems originate from normal close binaries with initial primary masses typically between 10 and 20 $M_\odot$. The system depicted here started with a 13 $M_\odot$ primary and a 6.5 $M_\odot$ secondary and an orbital period $P_o$ = 2.58 days. The primary star fills its Roche lobe only after it has left the main sequence, 12 million years after the birth of the system. The primary star then has a contracting helium core and is burning hydrogen in a shell around this core; its envelope is then expanding on a thermal timescale towards giant dimensions. The mass transfer which ensues is so-called stable "early Case B", in which case the mass lost by the primary star is captured by the secondary (e.g. see Pfahl et al. 2002). As soon as the primary overflows the Roche lobe, mass transfer to the 6.5 $M_\odot$ companion begins, causing the orbit to shrink until both stars have reached the same mass. After that the further mass transfer makes the orbit expand. The mass transfer continues until the primary star has lost its entire hydrogen-rich envelope and only its 2.5 $M_\odot$ helium core is left as a helium-burning helium star. At that time the orbital period has become 20.3 days and the companion now has a mass of 17 $M_\odot$. As the mass transfer proceeded by means of a disk (due to the orbital angular momentum of the transferred matter) this 17 $M_\odot$ star rotates extremely rapidly and therefore will be a B-emission star with an equatorial gas disk. Helium stars in the mass range between about 2.2 and 3.3 $M_\odot$ develop during their later evolution a degenerate O-Ne-Mg core, while their outer layers expand to giant dimensions (Habets 1986a,b; Dewi and Pols 2003). In a binary they will later on therefore go through a second, so-called

"case BB" phase of mass transfer by Roche-lobe overflow.

In the system of figure 2 this starts about 14.8 million years after the birth of the system and the star loses 0.3 $M_\odot$, leaving a 2.2 $M_\odot$ remnant in a 28.3day orbit. The core of this star shortly afterwards collapses to a neutron star due to electron-captures on the O-Ne-Mg ions. Assuming a neutron star with a gravitating mass of 1.4 $M_\odot$ to be left after this supernova explosion, and the supernova mass ejection to have been spherically symmetric, one finds that due to the explosive mass loss the orbital period has increased to 30.63 days and an orbital eccentricity of 0.043 has been induced while the system was imparted a runaway velocity of 14 km/sec.

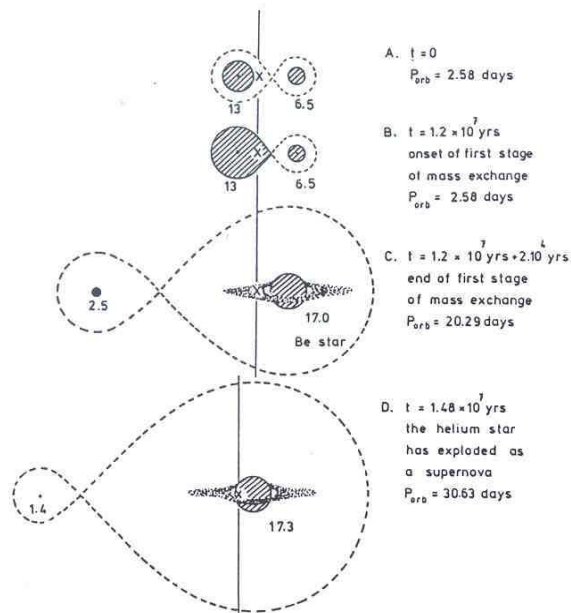

*Fig. 2    Model for the formation of a B-emission X-ray binary out of a close pair of B-type stars with masses of 13.0 and 6.5 $M_\odot$. The numbers in the figure indicate masses in solar masses (Habets, 1986a,b). The transferred matter, accreted by the companion, had high orbital angular momentum causing this star to become a very rapid rotator surrounded by a Be-star disk. Between the phases C and D the 2.5 $M_\odot$ helium star loses another 0.3 $M_\odot$ to its companion by Roche-lobe overflow (stellar wind mass loss was neglected).Its orbital period before the collapse was 28.3 days. Assuming aspherically symmetric supernova mass ejection the final orbital eccentricity is 0.043 and the runaway velocity of the system is 14 km/s. Further explanation in the text.*

**2.3. Evidence for a bimodal distribution of kick velocities imparted to neutron stars at birth.**

Figure 2 shows that with spherically symmetric supernova mass ejection one expects the orbital eccentricities of the B-emission X-ray binaries to be small. This holds for all primary masses in the range 10-20 $M_\odot$ and for all kinds of initial mass ratios of the systems. In view of the relatively low amounts of explosively ejected mass from the helium stars, always the eccentricity imparted by the spherically symmetric mass ejection from the helium star will be less than 0.2.

It will be clear therefore that in order to obtain the high orbital eccentricities of all Be/radio-pulsar systems (table 1) and of more than half of the Be/X-ray binaries, the neutron stars in these systems must have received a considerable kick velocity, of some 100 to 200 km/sec at their births (Pfahl et al. 2002). The fact, however, that 6 out of 13 Be/X-ray binaries have a low eccentricity, <0.25, then implies that in these systems the neutron stars received a kick velocity not larger than a few tens of km/sec, as first noticed by Pfahl et al. (2002). These authors showed that in order to obtain the division of the Be/X-ray binaries (plus Be/radio pulsar binaries) into a low-eccentricity group and a high-eccentricity group, the distribution of the kick velocities imparted to neutron stars at birth should be bi-modal, having one peak at v < 50 km/sec and a second peak at v ~ 100 - 200 km/sec. They suggested that if neutron stars originated from systems with stable early case B or C mass transfer, they only receive a small kick at birth, whereas in all other cases neutron stars would receive a large birth kick velocity. "Small" meaning here: a Maxwellian kick velocity distribution with characteristic velocity about 20 km/s and "large" meaning: such a distribution with characteristic velocity 100 km/s or more. They showed that since the low-eccentricity systems have in general much lower X-ray luminosities than the high-eccentricity systems, the galactic population of the low-eccentricity systems must be very large, of order of at least a few thousands.

**2.4. The double neutron stars: evolutionary products of B-emission X-ray binaries.**

Figure 3 depicts the later evolution of a wide B-emission X-ray binary. Once the B star evolves off the main sequence towards a red giant, the envelope of this star will at a certain moment engulf the neutron star, leading to the formation of a Common Envelope (CE) in which the neutron star and the helium core of the giant are embedded (cf. Taam, Bodenheimer and Ostriker 1978; Paczynski 1976; van den Heuvel 1976). One expects these two "stars" to spiral towards each other due to the large frictional drag they experience while orbiting inside this common envelope. If the frictional energy produced by this spiral in (augmented with released accretion energy) exceeds the gravitational binding energy of this envelope, the envelope will be driven off and a very close binary, composed of the helium core of the Be star plus the neutron star, will remain. The minimum condition for the energy release during spiral in to exceed the gravitational binding energy of the star's envelope is that the drop in system's gravitational binding energy between the original and final orbits should exceed the gravitational binding energy of the envelope (Webbink 1984). One finds that in order that this condition be fulfilled the orbital radius should shrink by at least a factor of a few hundred (Webbink, 1984; see for example Bhattacharya and van den Heuvel 1991, van den Heuvel 1994). In order to obtain then that in the final orbit the helium star does not overflow its Roche lobe already at the start of its evolution, the final orbital period should be at least a few hours, which implies that the initial Be/X-ray binary must have had an orbital period longer than about 100 days. In systems that have shorter orbital periods there is not sufficient orbital binding energy available to drive off the envelope and the neutron star will spiral all the way into the core of its companion, producing a "Thorne-Zytkow star"(Thorne and Zytkow 1977): a star with a neutron star in its center. This is expected to be the fate of all the standard High Mass X-ray Binaries, such as Cen X-3, SMC X-1, etc. The final outcome in the system of figure 3 is a very close binary composed of a 3 $M_\odot$ helium star and a neutron star and it was suggested that such a system would resemble the peculiar 4.8-hour orbit X-ray binary Cygnus X-3 (van den Heuvel and De Loore 1973). This was confirmed in 1992 by the discovery that the companion star of Cygnus X-3 is a Wolf-Rayet star, which means: a helium-burning helium star with a very strong stellar wind (van Kerkwijk et al. 1992).

The radius of a helium star more massive than 3.5 $M_\odot$ does not expand much during its further evolution and about 1 million years later (or less) its core collapses and the star explodes as a supernova. The outcome with spherically symmetric supernova mass ejection, which leaves a 1.4 $M_\odot$ neutron star remnant, will for $M_{He} < 4.2$ $M_\odot$ be a bound system consisting of two neutron stars in an eccentric orbit. A system with the orbital period of Cygnus X-3 (4.8 hours) will thus produce a double neutron star with an orbital period of order 7 or 8 hours, closely resembling the Hulse-Taylor binary radio pulsar system PSR B1913+16 (P=7.75 hours, e=0.61). If a kick velocity of a few hundred km/s is imparted to the neutron star at its birth, or if the mass of the helium star at the time of the explosion is over 4.2 $M_\odot$, the system may be disrupted, leading to the formation of two runaway neutron stars, one young and one old. Like in the double neutron star systems, the "old" one of the two neutron stars is a so-called "recycled" radio pulsar: it experienced a complex history, first: of accretion in a Be/X-ray binary system, subsequently: of spiral-in into the envelope of its companion star and finally: as the companion of the helium star in a very close binary for of order a one million years. Due to the accretion of mass with angular momentum its spin was accelerated, causing it to become a fast spinning pulsar (Bisnovatyi-Kogan and Komberg 1975); and the accretion process itself is expected to have caused a reduction of the strength of its surface dipole magnetic field (Van den Heuvel and Taam 1986, Ruderman 1999, Ruderman and Chen 1999). The Hulse-Taylor binary pulsar PSR 1913+16 is expected to be the outcome of this type of evolution (Flannery and van den Heuvel 1975, De Loore et al. 1975). This star has a $10^{10}$ Gauss dipole magnetic field strength, some two orders of magnitude weaker than the average for normal single pulsars, and an abnormally short rotation period (59 ms), indicating that this is the "old" neutron star in the system and has been recycled (Smarr and Blandford 1976; Radhakrishnan and Srinivasan 1982; Srinivasan and van den Heuvel 1982). Due to its weak magnetic field the electromagnetic spin down torque experienced by this recycled pulsar is much smaller than for strong-field pulsars, so it will have a very long spin-down timescale, of several hundreds of millions of years, whereas normal strong-field pulsars typically have spin down timescales of only a few tens of millions of years. The second neutron star born in the system did not undergo recycling and is therefore expected to resemble the normal single strong-magnetic-field pulsars. It therefore is expected to spin down rapidly and disappear within a few tens of millions of years into the pulsar "graveyard", where it is no longer observable. As its recycled companion is expected to remain observable for hundreds of millions of years (observational evidence indicates that neutron star magnetic fields do not decay spontaneously; Kulkarni 1986), this explains why in the Hulse-Taylor binary pulsar as well as in the other double neutron stars discovered since then (table 2), one so far only observed the weak-field recycled pulsar (Srinivasan and van den Heuvel 1982), simply because statistically this long-lived pulsar has a much higher probability to be observable than its short-lived companion.

Only very recently a double neutron star system was discovered in which both neutron stars are observed as pulsars: PSR J0737-3039AB. Indeed in this system the first-discovered pulsar is again the recycled one (P=23 ms, B=6x10$^9$G, Burgay et al.2003), but the second-discovered one is a normal strong-magnetic field pulsar with a 2.7 second period and a magnetic field strength of ±1.2x10$^{12}$ G (Lyne et al.2004), just as predicted from the binary recycling model for the formation for double neutron stars depicted in figure 3.

**2.5. The Double neutron star systems: evidence for low masses and low birth kick velocities of the second-born neutron stars**

Table 2 lists the observed parameters of the seven double neutron stars in the galactic disk that are presently known, together with the parameters of the 4.8 hour eccentric-orbit white dwarf-neutron star system PSR J1145-6545. The pulsar in the latter system is a young non-recycled pulsar with a strong magnetic field, just like the second-discovered neutron star in the double pulsar system PSR J0737-3039AB (Lyne et al. 2004).

These non-recycled strong-field pulsars are the second-born compact objects in their binary systems (see above).

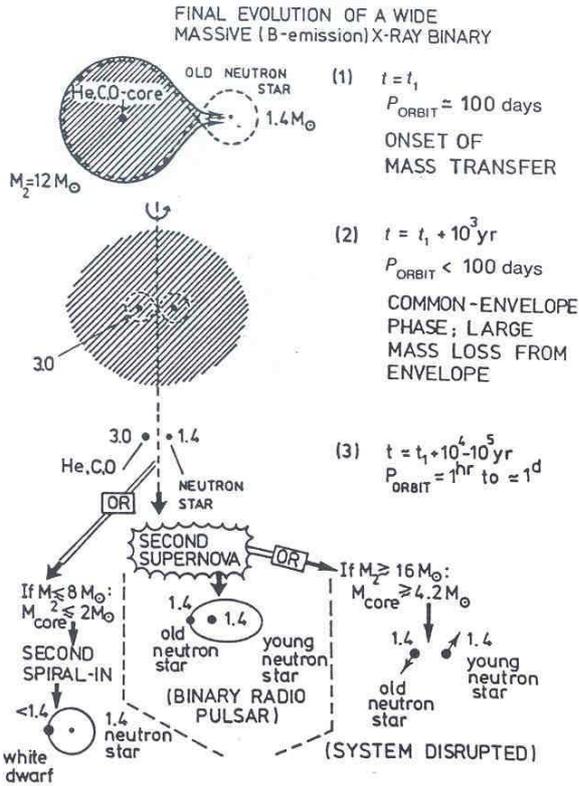

*Fig. 3 Various possibilities for the final evolution of a wide B-emission X-ray binary (orbital period ≥ 100 days). When the Be star has evolved into a giant, the neutron star is engulfed by its envelope, leading to the formation of a Common-Envelope system in which the neutron star spirals down towards the heavier-element core of the giant. After the ejection of the common envelope a very close binary remains, consisting of the neutron star and the helium (or heavier-element) core of the giant. If the helium core is less massive than 2.2 $M_\odot$ it will go through a further mass transfer and spiral-in phase and will leave a massive white dwarf star in a circular orbit around the neutron star. Cores in the mass range 2.2-3.3 $M_\odot$ also go through a further mass transfer phase, which is followed by core collapse and the formation of a second neutron star, in an eccentric orbit around the first one. If the mass of the core at the time of (symmetric) collapse is above 4.2 $M_\odot$, or if large velocity kicks are imparted at birth to the neutron star, the system may be disrupted in the second supernova explosion, leading to the formation of two runaway neutron stars.*

In that sense the white dwarf in the system of PSR J1145-6545 could be called a "recycled" white dwarf, just as the 23 ms pulsar PSR J0737-3039A is the recycled neutron star in that system. (We did not include in the table the double neutron star PSR B2127+11C in the globular cluster M15, as this system is expected to have formed by a completely different mechanism - by a multi-body interaction in the cluster core - than the systems in the disk). The table shows - as explained in the foregoing section - that with the exception of PSR J0737-3039B, the observed pulsars in the double neutron star systems are the recycled ones and not the last-born ones.

It is interesting to notice from the table that for three systems the mass of the second-born neutron star has been measured from the relativistic effects observed in these binary systems, and in all these three cases this mass comes out to be quite low: between 1.25 and 1.30 $M_\odot$:

- In the double pulsar system PSR J0737-3039 the doppler orbits of both stars have been measured, as well as the relativistic periastron precession and the Shapiro delay; this resulted in very accurately measured masses: for star A of 1.337(5) $M_\odot$ and for star B of 1.250(5) $M_\odot$.
- In the white dwarf-pulsar system PSR J1145-6545 the measurement of the general relativistic periastron precession rate of 5.308 degrees/yr yields a very accurate sum of the masses of the components of 2.288(3) $M_\odot$ and the accurate measurement of the Shapiro delay yields a mass of the white dwarf of 1.00(2) $M_\odot$, yielding a mass of the neutron star of 1.28(2) $M_\odot$ (Bailes 2004; in all cases the numbers within parentheses indicate the uncertainty in the last digit).
- The system PSR J1829+2456, a 41 msec pulsar in an 28-hour eccentric orbit, was discovered by Champion et al.(2004), who measured its general relativistic periastron precession rate to be 0.28(1) degrees per year, yielding a total system mass of 2.53(10) $M_\odot$. Using the maxium an minimum possible values of the periastron advance rate and assuming the pulsar to have a mass not smaller than the smallest accurately measured neutron star mass( 1.25 $M_\odot$), the authors constrain the masses of the pulsar and its companion to be located within a triangle in the pulsar-mass vs. companion-mass diagram. The location of the center of this triangle yields a most likely companion mass of 1.27 (+.11,-.07) $M_\odot$ and pulsar mass 1.30 (+.05, -.05) $M_\odot$ (see figure 3 of Champion et al. 2004).

Furthermore there are three systems in which the sum of the masses is known from the measured relativistic rate of periastron precession, and for which in two cases the observations indicate that the companion of the pulsar (i.e.: the second-born neutron star) has the lower mass of the two, not larger than 1.30 $M_\odot$:

- In the eccentric-orbit double neutron star system PSR J1518+4904 the most likely mass of the pulsar is the largest of the two and the companion has the lower mass of the two (Nice et al.1996). As the sum of the masses is only 2.62 $M_\odot$, it is then most likely that the companion has a mass not larger than 1.30 $M_\odot$. [The measurement of the relativistic periastron precession rate of 0.0111(2) degrees per year by Nice et al. (1996) yields a total system mass of 2.62(7) $M_\odot$; in combination with the mass function of 0.115988 $M_\odot$, this leads to the most likely masses of the pulsar and its companion of 1.56 (+.13, -.44) $M_\odot$ and 1.05 (+.45,-.11) $M_\odot$, respectively (Thorsett and Chakrabarty 1999)].

- The second "long" orbital period eccentric system PSR J1811-1736 has a pulse period of 104 ms and an orbital period of 18.8 days. The best estimates of the companion mass and pulsar mass are: 1.11 (+.53, -.15) $M_\odot$ and 1.62 (+.22, -.55) $M_\odot$, respectively (Stairs 2004), while the total system mass derived from the relativistic periastron precession rate is 2.60 $M_\odot$ (Lyne et al.2000). Also here the fact that the companion is most likely to have the lowest mass of the two indicates its mass to be below 1.30 $M_\odot$.

- Finally, Lyne (2004) reported the discovery of the eccentric double neutron star system PSR J1756-2251, which has a pulse period of 28.5ms and a binary period of 7.7 hours (almost identical to that of the Hulse-Taylor pulsar), but an orbital eccentricity of only 0.18. Like all the others it clearly is a recycled pulsar and from the measurement of its relativistic periastron precession the total system mass is estimated to be 2.60(2) $M_\odot$, i.e.: again relatively small. As yet no estimates of the component masses are available but it will be clear that at least one of the components will have a mass not larger than 1.30 $M_\odot$.

The first conclusion from the above and from the table is that in 5 out of the 7 double neutron stars in the galactic disk the sum of the masses is relatively small, namely ~ 2.60 $M_\odot$, implying that at least one of the neutron stars has a mass not larger than 1.30 $M_\odot$. The above summary shows that in at least 4 of these 5 cases it is the "companion", i.e.: the second-born neutron star that has the smaller mass of the two and therefore has: M < 1.30 $M_\odot$. The same holds for the neutron star in the white dwarf-neutron star binary PSR J1145-6545, which also is the second-born compact object in that system. Furthermore we see that - including the white-dwarf system - in five out of these 6 systems with low-mass second-born neutron stars the orbital eccentricity is quite small: in the range from 0.088 for PSR J0737-3039 to 0.249 for J1518+4904. The only exception is J1811-1736, which has e=0.828. Among the two remaining systems, which have a sum of the masses substantially larger than 2.60 $M_\odot$, also one, the Wolszczan pulsar PSR B1534+12 has a relatively low orbital eccentricity, of only 0.274.

One thus notices that despite the fact that it is generally assumed that pulsars receive large kick velocities at birth, in the range of several hundreds of km/s, velocities that will make the orbits very much more eccentric than 0.25, in 6 out of these 8 systems the supernova explosion that produced the second-born neutron star in the system, did NOT induce a large orbital eccentricity! This indicates that in six out of these eight cases the neutron stars apparently did belong to the low-kick portion of the bimodal kick velocity distribution that is required to explain the presence of a sizeable population of B-emission X-ray binaries with low orbital eccentricities (Pfahl et al. 2002; see section 2.3, above).

Combining this with the above made observation that in the second supernova explosion in general neutron stars with low masses have been produced, in the range 1.25 - 1.30 $M_\odot$, one is strongly tempted to conclude that apparently in the formation of a low-mass neutron star no large kick velocity is imparted. Notice that for the B-emission X-ray binaries with low orbital eccentricities no such conclusion could be drawn since here the masses of the neutron stars are not known.

### 2.6. Nature of the low-kick neutron stars: formation by electron-capture collapse of a degenerate O-Ne-Mg core?

For a neutron star a gravitational mass of 1.25 -1.30 $M_\odot$ corresponds to a baryonic mass of about the Chandrasekhar mass of 1.44 $M_\odot$. The loss of gravitational binding energy in the formation of the

*Table 2 Double neutron star binaries and the eccentric-orbit white-dwarf neutron star system J1145-6545*

| Pulsar Name | Spin Per. (ms) | $P_{orb}$ (d) | e | Compan. Mass ($M_\odot$) | Pulsar Mass ($M_\odot$) | Sum of masses ($M_\odot$) | Ref |
|---|---|---|---|---|---|---|---|
| J0737-3039A | 22.7 | 0.10 | 0.088 | 1.250(5) | 1.337(5) | 2.588(3) | (1) |
| J0737-3039B | 2770 | 0.10 | 0.088 | 1.337(5) | 1.250(5) | 2.588(3) | (1) |
| J1518+4904 | 40.9 | 8.63 | 0.249 | 1.05 (+0.45) (-0.11) | 1.56 (+0.13) (-0.45) | 2.62(7) | (2) |
| B1534+12 | 37.9 | 0.42 | 0.274 | 1.3452(10) | 1.3332(10) | 2.678(1) | (3) |
| J1756-2251 | 28.5 | 0.32 | 0.18 | - | - | 2.60(2) | (4) |
| J1811-1736 | 104 | 18.8 | 0.828 | 1.11 (+0.53) (-0.15) | 1.62 (+0.22) (-0.55) | 2.60 | (3) |
| J1829+2456 | 41.0 | 1.18 | 0.139 | 1.27 (+0.11) (-0.07) | 1.30 (+0.05) (-0.05) | 2.53(10) | (5) |
| B1913+16 | 59 | 0.33 | 0.617 | 1.3873(3) | 1.4408(3) | 2.8281(1) | (3) |
| J1145-6545 | 394 | 0.20 | 0.172 | 1.00(2) | 1.28(2) | 2.288(3) | (6) |

References: (1) Lyne et al. (2004); (2) Nice et al. (1996); (3) Stairs (2004); (4) Lyne (2004); (5) Champion et al. (2004); (6) Bailes (2004).

neutron star is, depending on the assumed equation of state, in the range $(0.11 - 0.13)M_b c^2$, which for $M_b$=1.44 $M_\odot$ corresponds to an "effective" loss of gravitational mass in the core collapse of between 0.16 $M_\odot$ and 0.19 $M_\odot$. This would result in neutron stars with a gravitational mass of about 1.25 to 1.28 $M_\odot$.

It is well known that there are two ways in which a neutron star can be formed (cf. Nomoto 1982; Miyaji et al. 1980):

(1) By the electron-capture collapse of a degenerate O-Ne-Mg core when the mass of this core approaches the Chandrasekhar limit. This is expected to happen in single hydrogen-rich stars with initial masses in the range 8 to about 10 $M_\odot$ and in helium stars in the mass range about 2.2 to 3.3 $M_\odot$ (Dewi and Pols 2003 and references therein). In binaries the latter helium star masses correspond to initial stellar masses in the range about 9 to 14 $M_\odot$ for so-called "case B" mass transfer.

(2) By the photo-disintegration collapse of an Fe-core which marks the endpoint of all nuclear burning stages in the star. This will occur in single stars more massive than about 10 $M_\odot$ and in helium stars more massive than about 3.3 $M_\odot$. Evolutionary calculations by Timmes et al. (1996) show that in single hydrogen-rich stars with masses below 19 $M_\odot$ the collapsing Fe core has a mass of about 1.3 $M_\odot$ and for initial masses larger than 19 $M_\odot$ it has a mass of about 1.7 $M_\odot$. Taking into account some fall-back of matter and the loss of gravitational binding energy, the Fe-cores in the first mass range are expected to produce neutron stars of about 1.3 to 1.5 $M_\odot$ and in the second mass range: neutron stars with masses probably upwards from about 1.7 $M_\odot$ (and also: black holes).

It should be noticed here that there is one neutron star known with a most likely mass of about 1.85 $M_\odot$, which might be the product of the collapse of a star that started out more massive than 19 $M_\odot$: the X-ray pulsar Vela X-1 (Barziv et al. 2001).

The masses of the second-born neutron stars in the low-eccentricity binary pulsar systems listed in table 2 strongly suggest that these neutron stars originated from the electron-capture collapse of a degenerate O-Ne-Mg core. Let us examine the implications of this model. If these neutron stars indeed originated from electron-capture collapse, then we must be dealing with the products of helium stars in the mass range 2.2 - 3.3 $M_\odot$ where e-capture collapse is expected to occur.

During the later evolution of these helium stars their envelopes expand to giant dimensions during the phase of helium shell burning (see section 2.2). This will in a close binary cause further mass transfer from the helium star towards its companion by Roche-lobe overflow, before the final supernova collapse (just like in the model for the origin of a Be/X-ray binary, see figure 2). The helium star in a close helium star plus neutron star system will therefore further transfer mass to the first-born neutron star in the system. This mass transfer by Roche-lobe overflow can be stable for several tens of thousands of years (Dewi and Pols 2003; Dewi, 2003; Ivanova et al. 2003), which provides an ideal situation for spinning up the rotation of this first-born neutron star to a short period. During this Roche-lobe overflow these helium stars may easily lose some 0.5 $M_\odot$ (most of this mass will be lost from the system, Dewi and Pols 2003; Ivanova et al. 2003). Furthermore, these helium stars also have non-negligible stellar wind mass loss during their entire evolution. Therefore, at the time of their core collapse the masses of these helium stars have come down to values typically in the range 1.6 to 2 $M_\odot$ (Pols and Dewi 2003; Ivanova et al. 2003), which implies that the mass loss in the final supernova explosion is very small (~0.15 to ~0.55 $M_\odot$). As to the origins of the kick velocities imparted to neutron stars at birth the present thinking is that these are related to the low-mode convective motions that develop in the regions surrounding the collapsing core if the neutrino luminosity of the core is sufficiently large to drive a supernova explosion (Janka and Mueller, 1994: Fryer and Heger, 2000). This convection can develop from random motions behind the SN shock and will cause the ejecta to be globally asymmetric. Scheck at al. (2004) find that in this case the neutron star can be accelerated on a timescale of order a second to a velocity as large as 500 km/s.

It is clear, however, that for this process to work there must be sufficient mass in the ejecta. If the mass of the ejecta is as tiny as we expect it to be in the case of the exploding remnants of the helium stars undergoing electron-capture collapse in helium-star plus neutron star binaries (or helium star plus Be binaries) the induced kick velocities may be expected to be relatively small (Pfahl et al. 2002; Fryer 2004).

An alternative hypothesis might be that the electron-capture collapse of a degenerate O-Ne-Mg-core is an intrinsically much more symmetric event than the photodisintegration collapse of a Fe core, i.e.: that only in the case of a collapsing Fe core a large kick velocity is imparted to the neutron star, but not in the case of the e-capture collapse of an O-Ne-Mg core. This alternative hypothesis is also consistent with the observational facts. It would automatically produce a bimodal distribution of the kick velocities of radio pulsars. On the basis of the shape of the Initial Mass Function one would expect roughly half of the Be/X-ray binaries to originate from systems with primary stars in the mass range 8 to 14 $M_\odot$, where the helium stars, which evolve towards an electron-capture collapse, are being produced by case B mass transfer.

The higher-eccentricity Be/X-ray binaries would then have originated from systems in which the initial mass of the primary star was > 14 $M_\odot$ and the double neutron stars with high orbital eccentricities (PSR B1913+16 and PSR J1811-1736) from helium star plus neutron star systems in which the helium star was more massive than about 3.3 $M_\odot$ at the end of the common-envelope evolution (at the onset of the helium burning).

## 2.7. Discussion and conclusions: three kinds of neutron stars?

The fact that 5 out of the 7 double neutron stars in the Galactic disk and also the white dwarf plus young neutron star binary PSR J1145-6545, have small orbital eccentricities (between 0.088 and 0.274), indicates that in 6 out of 8 cases the velocity kick imparted at birth to the second-born neutron star was small. [If random kick velocities in the range 100 to 200 km/s, corresponding to the high-velocity peak of the kick-velocity distribution (Pfahl et a.2002) had been imparted to all these neutron stars at birth, such a clustering at low orbital eccentricities would be highly unlikely.] Thus, the second-born neutron stars in 6 out of these 8 systems appear to belong to the "low kick" population of the bi-modal kick velocity distribution, just like the neutron stars in the low-eccentricity B-emission X-ray binaries. This fact, in combination with the observation that (at least) in 5 out of the 6 systems with low eccentricities the neutron stars have a low mass, in the range 1.25 to 1.30 $M_\odot$, strongly suggests that in general low-mass neutron stars do not receive a large kick velocity at birth.

These facts, in combination with the presence of also a "high kick" neutron star population, some of which are known to have a relatively high mass (the companion of the high-eccentricity Hulse-Taylor binary pulsar has a mass of 1.38 $M_\odot$) may find a consistent explanation if one assumes that there are two fundamentally different categories of neutron stars, formed by different mechanisms, viz.:

1). A population of low-mass neutron stars which received a low kick velocity at birth and were formed by the electron-capture collapse of a degenerate O-Ne-Mg core. These neutron stars originate in binaries from stars that initially had masses < 14 $M_\odot$ (for case B evolution) or <12 $M_\odot$ (for case C evolution).

2). A population of more massive neutron stars which formed by the photo-disintegration collapse of an iron core and received a large kick velocity at birth. These neutron stars originate from stars in binaries with initial masses >14 $M_\odot$ (case B evolution) or >12 $M_\odot$ (case C evolution). The latter population is expected to be further subdivided into two classes: those originating from stars with initial masses <19 $M_\odot$ would leave neutron stars with masses ~1.4 $M_\odot$, those from stars with initial masses >19 $M_\odot$ leaving neutron stars more massive than about 1.7 $M_\odot$. Vela X-1, with its mass of 1.85 $M_\odot$ has been suggested to belong to this latter class. The above would imply that according to their formation mechanisms, there are in fact three categories of neutron stars.

The fact that there are so few high-eccentricity double neutron stars may, in terms of the above described formation models, simply be explained as due to several selection effects. In the first place the velocities of the "high-kick" part of the kick velocity distribution are quite high, ranging to over 500 km/s (Hansen and Phinney 1997). Such kicks will rather easily disrupt the binary systems in the first or the second supernova explosion, whereas of the systems with "low-kick" second-born neutron stars, none will be disrupted in the second supernova explosion. Secondly, as the more massive neutron stars originate from more massive helium stars, also the mass ejected in the second supernova will be systematically larger in these systems, which adds to their disruption probability. In the third place there is the fact that due to the shape of the initial mass function, lower-mass stars are more numerous than massive ones. The fact that a fraction of the more massive systems may already be disrupted in the first supernova explosion implies that the population of the (lower mass) low-eccentricity Be/X-ray binaries is larger than that of the high-eccentricity systems, which favors the low-eccentricity Be/X-ray binaries as progenitors of double neutron stars.

These four effects combined may easily explain why the low-eccentricity systems, which originate from the lower-mass end of the B-emission X-ray binary population, dominate the double neutron star population. The "overabundance" of low-mass low-eccentricity Be/X-ray binaries among the progenitors of systems undergoing a second supernova explosion may explain why there are so few single recycled pulsars - as hardly any of these systems will be disrupted in the second supernova explosion.

**Acknowledgements**: I thank D. Manchester, D. Lorimer, A. Lyne, P. Podsiadlowski, J.D.M. Dewi, B.W. Stappers and other participants of the 2004 Aspen Winter workshop on pulsars for many insightful discussions.